\documentclass{mem}
\usepackage{natbib}
\usepackage{txfonts}
\usepackage{balance}
\usepackage{graphicx}
\usepackage[a4paper,breaklinks,dvipdfm]{hyperref}
\idline{84}{1}
\usepackage{txfonts} 

\begin{document}
\def\teff{$T\rm_{eff }$}
\def\kms{$\mathrm {km s}^{-1}$}

\title{
The relationship between X-ray emission and accretion in X-ray selected AGNs
}

\subtitle{}

\author{
R. Fanali,\inst{1,2} 
A. Caccianiga,\inst{1}
P. Severgnini,\inst{1}
R. Della Ceca,\inst{1}
M. Dotti,\inst{2}
E. Marchese,\inst{2,1}
A. Corral,\inst{3}
          }

\offprints{rossella.fanali@brera.inaf.it}

\institute{
INAF - Osservatorio Astronomico di Brera, Milan, Italy
\and
Universit\'a Degli Studi di Milano Bicocca, Milan, Italy
\and
National Observatory of Athens (NOA), Athens, Greece
}

\authorrunning{R. Fanali et al.}

\titlerunning{X-ray emission and accretion rate
}

\abstract{We study the link between the X-ray emission in radio-quiet AGNs and the accretion rate on the central Supermassive Black-Hole (SMBH) using a well-defined and statistically complete sample of $70$ type$1$ AGNs extracted from the XMM-Newton Bright Serendipitous survey (XBS). To this end, we search and quantify the statistical correlations between the main parameters that characterize the X-ray emission (i. e. the X-ray spectral slope and the X-ray “loudness”), and the accretion rate, both absolute ($\dot{M}$) and relative to the Eddington limit (Eddington ratio, $\lambda$). Here, we summarize and discuss the main statistical correlations found and their possible implications on current disk-corona models.
}

\maketitle

It is now accepted that the engine of AGNs is powered by the accretion of matter onto the SMBH, placed in the center of the host galaxy: the matter is heated ($\sim 10^6$ K) through viscous and magnetic process and forms an accretion disk around the SMBH emitting in the UV-optical region. A fraction of energy is also emitted in the X-ray band with a spectrum that can be represented, at a zeroth order, by a power-law from 0.1 to 100 keV. It is believed that X-rays are produced in a hot corona ($\sim 10^8$ K) reprocessing the primary UV-optical emission of
the disk via inverse-Compton mechanism (\citealp{H}).
The main properties of X-ray emission change significantly from source to source. Recent results suggest that the differences can be partly related to the value of accretion rate or to the black-hole mass (\citealp{Grupe et al. 2010}, \citealp{Risaliti et al. 2009}, \citealp{V}).  
Understanding if and how the X-ray properties are related to accretion is a fundamental step to study the link between disk and corona, i. e. the two main components of an AGN.

The aim of this work is to establish the actual link between X-ray properties and the parameters that quantify the accretion rate by analyzing a well defined sample of $70$ type$1$ AGNs selected from the XMM-Newton Bright Serendipitous survey (XBS). 
In this work we study the spectral index $\Gamma$ between 0.5 and 10 keV and the bolometric correction $K_{\rm bol}$, defined as the ratio between bolometric luminosity and $2-10$ keV luminosity. $\Gamma$ gives direct information about the energy distribution of the electrons in the corona, while $K_{\rm bol}$ quantifies the relative importance between disk and corona. The approach followed is to search for statistically significant correlations between these parameters and the value of accretion rate, absolute ($\dot{M}$) and normalized to Eddington limit ($\lambda$), presented in (\citealp{Caccianiga et al. 2012}). 

\noindent{\bfseries Results}

\noindent We found that $\Gamma$ depends significantly on $\lambda$ while the dependence on $\dot{M}$ is weak. By using partial correlation we demonstrate that the observed correlations are not induced by redshift. We have found a similar $\Gamma-\lambda$ correlation also using the hard $2-10$ keV spectral index. Therefore the observed correlation is directly linked to the primary component of the X-ray emission and not due to secondary spectral component (e.g. soft excess). We also found a correlation between $K_{bol}$ and $\lambda$ while, again, the dependece on $\dot{M}$ is weak. 

\noindent Using the partial correlation analysis we have also verified that other correlations observed in our sample ($\Gamma-M_{BH}$, $\Gamma-FWHM(H_\beta)$, $K_{bol}-L_{UV}$) are probably just secondary correlations induced by the $\Gamma-\lambda$ and $K_{bol}-\lambda$ correlations.

\begin{figure}
\centering
\includegraphics[width=6.4cm]{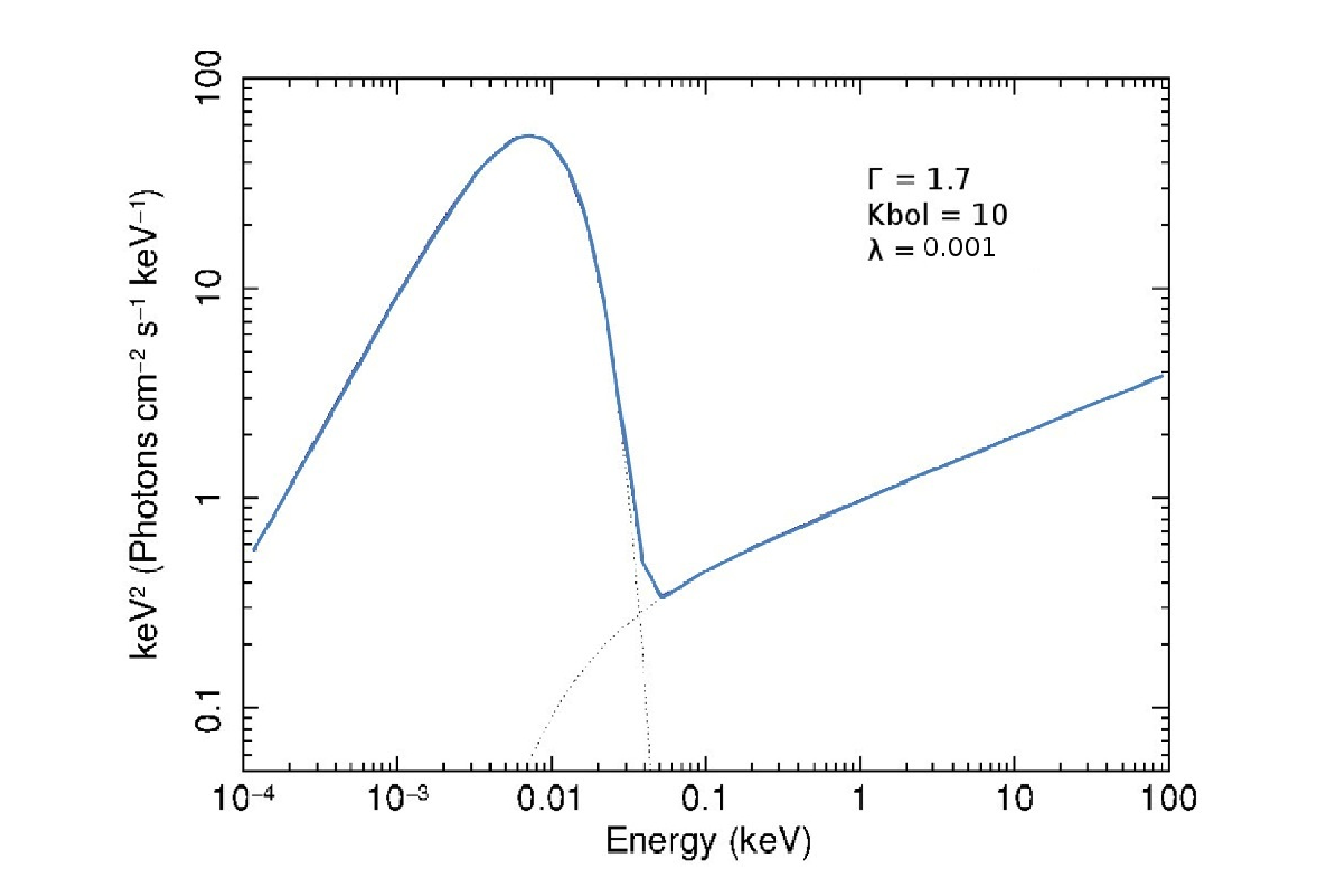}
\includegraphics[width=6.2cm]{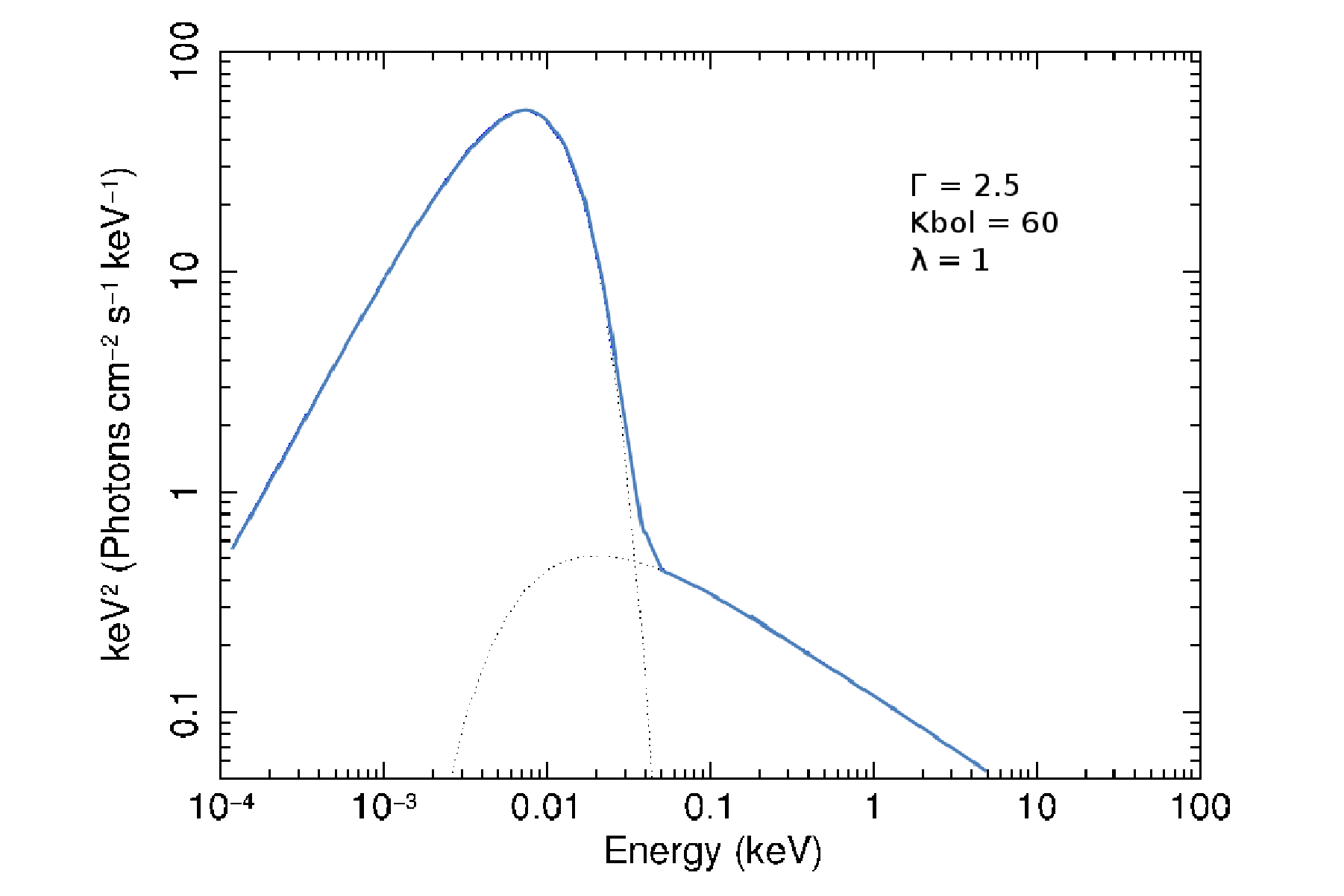}
\caption{Theoretical spectral energy distributions that represent two extreme cases of accretion: $\lambda \sim 10^{-3}$ (upper panel) and $\lambda \sim 1$ (lower panel). See text for more details.}
\label{fig:figure}
\end{figure}

We represent graphically the results of the $\Gamma-\lambda$ and $K_{bol}-\lambda$ correlations by showing the theoretical SEDs in two extreme cases of low and high Eddington ratios (Fig. \ref{fig:figure}): these SEDs were built using a Shakura $\&$ Sunyaev disk model plus a power-law in the range between $0.01$ and $100$ keV with a cut-off at $0.1$ keV. The values of $\Gamma$ and $K_{bol}$ are taken from our fits of the $\Gamma-\lambda$ and $K_{bol}-\lambda$ correlations.
To simplify the comparison, we assumed the same disk emission normalization in both cases. 
The comparison of the two SEDs suggests that the variation of $K_{\rm bol}$ could be entirely attributed to the variation of $\Gamma$. We have also tested this idea by using the partial correlation analysis and we conclude that the correlations observed in this work can be in principle explained only by the $\Gamma - \lambda$ correlation. This result suggests a speculative interpretation based on the electron cooling of corona: for high values of $\lambda$, a large number of photons comes from the accretion disk and cools corona electrons rapidly, thus poducing steep X-ray spectra and high values of $K_{\rm bol}$. For low $\lambda$, less photons are available and this makes electron cooling inefficient, thus producing flat X-ray spectra and low values of $K_{\rm bol}$.

%\begin{figure}
%\centering
%\subfigure
%{\includegraphics[width=7.0cm]{flat22.pdf}}
%\hspace{5mm}
%\subfigure
%{\includegraphics[width=6.7cm]{steep2.pdf}}
%\caption{Theoretical spectral energy distributions that represent two extreme cases of accretion: $\lambda \sim 10^{-3}$ (upper panel) and $\lambda \sim 1$ (lower panel). See text for more details.}
%\label{fig:figure}
%\end{figure}

In conclusion, this work shows that the X-ray properties depend on how much the central black hole is accreting with respect to the Eddington limit. Full details will be reported in Fanali et al. in prep.

\begin{acknowledgements}
We acknowledge Francesco Haardt for the useful discussions and ASI (grant n. I/008/06/0) for financial support.
\end{acknowledgements}

\bibliographystyle{aa}

\end{document}